\documentclass[aps,pra,reprint,groupedaddress,showkeys]{revtex4-1}
\usepackage{amsmath,graphicx,hyperref,verbatimbox,array,float}
\usepackage{indentfirst}
\usepackage{braket}
\usepackage{diagbox}
\usepackage{amssymb}
\usepackage{bbold}
\usepackage{colortbl}
\usepackage{multirow}
\usepackage{subcaption}
\newcommand{\figref}[1]{Fig.~\ref{#1}}

\renewcommand{\eqref}[1]{Eq.~(\ref{#1})}
\newcommand{\tr}{{\mathrm{Tr}}}

\begin{document}
\title{Quantum State Tomography of Qutrits by Single-Photon Counting with Imperfect Measurements}
\author{Jakub Szlachetka and Artur Czerwinski}
\email{aczerwin@umk.pl}
\affiliation{Institute of Physics, Faculty of Physics, Astronomy and Informatics \\ Nicolaus Copernicus University, Grudziadzka 5, 87--100 Torun, Poland}

\begin{abstract}
In this article, we introduce a framework for quantum state tomography of qutrits by projective measurements. The framework is based on photon-counting with measurement results distorted due to the Poisson noise and dark counts. Two different measurement schemes are investigated numerically and compared in terms of their efficiency for distinct numbers of photons per measurement. The accuracy of state reconstruction is quantified by figures of merit which are presented on graphs versus the amount of noise.
\end{abstract}
\keywords{quantum state tomography, photonic tomography, single-photon counting, qutrits}
\maketitle

\section{Introduction}
Quantum state tomography (QST) aims at determining the accurate representation of a physical system. This can be done by repeating various measurements, provided the source continues to produce identical copies of an unknown quantum state. The ability to efficiently achieve the complete knowledge of a quantum system is relevant to the development of modern quantum technologies \cite{Horodecki2021}.

In particular, photonic tomography relies on the ability to measure a specific number of counts for each type of measurement \cite{Altepeter2005}. The obtained photon counts are used to determine the probabilities corresponding to selected measurement settings. From a mathematical point of view, the quantum state that best describes the observed system can be derived from the probability formula, which is called the Born's rule \cite{Born1955}.

In general, any quantum measurement is described by a collection of operators $\{Z_{m}\}$. Assuming we measure quantum state $ \ket{\Psi}\in\mathcal{H} $, the probability of the outcome $m$ is given as:
\begin{gather}
    p(m) = \bra{\Psi} Z_{m}^{\dagger} Z_{m} \ket{\Psi} .
\end{gather}
Let us introduce:
\begin{gather}
    M_{m} \equiv Z_{m}^{\dagger} Z_{m} ,
\end{gather} 
which is a positive operator. Since the probabilities should sum to one, we have $\sum_{m} M_{m} = \mathbb{I}$, where $\mathbb{I}$ denotes the identity operator. A set of such measurement operators  $\{ M_{m} \}$ is referred to as a positive operator-valued measure (POVM) \cite{Nielsen2000}. If a POVM comprises such measurement operators $\{ M_{m} \}$ that are sufficient for state reconstruction, it is called informationally complete POVM (IC-POVM) \cite{Busch1991}.

Usually, special attention is paid to a particular class of POVMs that is called a symmetric, informationally complete, positive operator-valued measure (SIC-POVM) \cite{Renes2004}. Originally, SIC-POVMs are constructed from rank-one projectors, but their general properties have also been excessively studied \cite{Rastegin2014}. The key property of SIC-POVMs relates to their maximum efficiency in QST, i.e., such measurement schemes can be called optimal as far as the number of distinct measurement setups is considered.

On the other hand, mutually unbiased bases (MUBs) can be employed as an overcomplete measurement scheme \cite{Wootters1989,Durt2010}. Such measurement operators are believed to perform better under noisy measurements.

These types of measurements are commonly implemented in experiments involving photons. An example of such an experiment is a measurement of quantum interference of a three-photon state \cite{Menssen2017}. Moreover, it was experimentally possible to teleport a photonic qutrit \cite{Luo2019}. Furthermore, it was demonstrated empirically that high-dimensional quantum states, such as qutrits, could facilitate quantum key distribution (QKD) schemes \cite{Bogdanov2004,Groblacher2006}. Qutrits can also be useful in quantum computing, see, e.g. Refs.~ \cite{Klimov2003,Mc2005,Baekkegaard2019}.

The abundance of experiments prompts us to investigate QST of qutrits. Three-level states can be realized on photons by exploiting different degrees of freedom, for example, spatial \cite{Taguchi2009} or temporal \cite{SedziakKacprowicz2020}, which justifies the need for efficient tomographic techniques. In our approach, we focus on projective measurements of photons with results disturbed by dark counts and the Poisson noise. Similar problems were considered on qubits \cite{Czerwinski2021} and ququads \cite{Czerwinski2020}. However, in the above works, the task was tackled in the context of the frame theory, whereas, in the present manuscript, we analyze the problem within the quantum mechanical formalism. The main goal of this work is to compare two sets of measurements, i.e., a SIC-POVM and MUBs, in terms of their performance in QST of qutrits with noisy data, involving different numbers of photons per measurement.

In Sec.~\ref{framework}, we present the framework of quantum tomography for a general qutrit state along with assumptions concerning the noise introduced into the measurements and figures of merit to quantify the accuracy of state reconstruction. Then, in Sec.~\ref{results}, the results are introduced and analyzed. We investigate qutrit state reconstruction with two distinct POVMs -- one defined by the elements from the MUBs \cite{Kurzynski2016} and the other involving the SIC-POVM \cite{PaivaSanchez2010}. The figures of merit, which allow one to compare the two sets of measurement operators, are presented graphically with error bars. The plots present the average fidelity, purity, and entropy for two selected numbers of photons versus the amount of dark counts. The article is summarized by a discussion in Sec.~\ref{discussion}.

\section{Framework for state reconstruction}\label{framework}

\subsection{Methods}\label{methods}
The framework for qutrit tomography is based on single-photon counting with measurements influenced by the Poisson noise and dark counts. We assume that our source generates photons, prepared in the same quantum state described by a vector from the $3-$dimensional Hilbert space. If $M_k$ stands for an arbitrary measurement operator ($M_k \geq 0$) and $\mathcal{N}$ denotes an average number of photons per measurement, we obtain a formula for the expected photon count:
\begin{equation}\label{met1}
    n^{E}_k := \mathcal{N} \, \tr \left(M_k \rho \right),
\end{equation}
where $\rho$ is a density matrix which represents an unknown state of the photon. Since we assume that the experimenter has no knowledge about the state in question, we follow the Cholesky decomposition \cite{James2001,Altepeter2005}:
\begin{equation}\label{met2}
    \rho = \frac{T^{\dagger} T}{ \tr \left( T^{\dagger} T \right)},
\end{equation}
where $T$ denotes a lower triangular matrix, which in the case of qutrits takes the form:
\begin{equation}\label{met3}
	T =\begin{pmatrix} t_1 & 0 & 0 \\ t_4 + i \,t_5 &  t_2 & 0 \\  t_8 + i \,t_9 & t_6 + i \,t_7 & t_3 \end{pmatrix}.
\end{equation}
Thus, the problem of state reconstruction for qutrits can be translated into finding the real numbers: $t_1, t_2, \dots, t_9$. The Cholesky factorization guarantees that the result of the QST framework must be physical, i.e., $\rho$ is positive semidefinite, Hermitian, of trace one.

The measured photon counts, $n^{M}_k$, are generated numerically by utilizing a standard parametrization of a pure qutrit state:
\begin{equation}\label{met4}
    \ket{\psi_{in}} = \begin{pmatrix} \cos \frac{\theta}{2} \, \sin \frac{\delta}{2} \\\\ \sin \frac{\theta}{2} \, \sin \frac{\delta}{2} \,e^{i \phi_{12}}  \\\\  \cos \frac{\delta}{2} \,e^{i \phi_{13}}  \end{pmatrix},
\end{equation}
where $\theta, \delta \in [0, \pi]$ and $\phi_{12}, \phi_{13} \in [0, 2 \pi)$. We select a sample of $5\,184$ qutrits such that the parameters cover the full range. Then, in order to make the framework realistic, we impose dark counts, which means that the detector, apart from the intended state produced by the source, receives a background noise modeled by the maximally mixed state. Thus, the received state takes the form:
\begin{equation}\label{met5}
    \rho_{in} = (1-p) \ket{\psi_{in}} \! \bra{\psi_{in}} + \frac{p}{3} \mathbb{1}_3,
\end{equation}
where $\mathbb{I}_3$ stands for the $3 \times 3$ identity matrix and $p \in [0,1]$ is referred to as the dark count rate. Next, we can introduce a formula for the measured photon counts:
\begin{equation}\label{met6}
     n^{M}_k = \mathcal{N}_k \tr \left(M_k \rho_{in} \right),
\end{equation}
where $\mathcal{N}_k$ denotes a number generated randomly for each measurement operator from the Poisson distribution characterized by the mean value $\mathcal{N}$. In this way, we impose the Poisson noise \cite{Hasinoff2014}, which a typical source of errors in protocols based on photon-counting, see also Refs.~\cite{SedziakKacprowicz2020,Czerwinski2021}.

Finally, in order to estimate the unknown state \eqref{met2}, we apply the method of least squares and search for the minimum value of the function:
\begin{equation}\label{met7}
    f_{LS} (t_1, t_2, \dots, t_9)  = \sum_{k=1}^{\kappa} \left( n^{E}_k - n^{M}_k \right)^2,
\end{equation}
where $\kappa$ gives the number of measurement operators involved in the scenario. By minimizing the function \eqref{met7} we determine the parameters $\{t_1, \dots, t_9\}$ that fit optimally to the noisy measurements.

\subsection{Definition of measurements}\label{measurements}
Projective measurements implemented in the framework are defined by two sets of operators: the SIC-POVM and MUBs. For $\dim \mathcal{H} =3$, the SIC-POVM is generated by means of nine vectors \cite{PaivaSanchez2010}:
\begin{equation}\label{sicpovm}
\begin{split}
&\ket{\nu_0 ^0} = \frac{1}{\sqrt{2}} (\ket{0} + \ket{1}), \hspace{0.5cm} \ket{\nu_1 ^0} = \frac{1}{\sqrt{2}} (\overline{\eta} \ket{0} +\eta \ket{1}), \\
&\ket{\nu_2 ^0} = \frac{1}{\sqrt{2}} (\eta \ket{0} +\overline{\eta} \ket{1}),\\
&\ket{\nu_0 ^1} = \frac{1}{\sqrt{2}} (\ket{1} + \ket{2}), \hspace{0.5cm} \ket{\nu_1 ^1} = \frac{1}{\sqrt{2}} (\overline{\eta} \ket{1} +\eta \ket{2}),\\
&\ket{\nu_2 ^1} = \frac{1}{\sqrt{2}} (\eta \ket{1} +\overline{\eta} \ket{2}),\\
&\ket{\nu_0 ^2} = \frac{1}{\sqrt{2}} (\ket{0} + \ket{2}), \hspace{0.5cm} \ket{\nu_1 ^2} = \frac{1}{\sqrt{2}} (\eta \ket{0} +\overline{\eta} \ket{2}),\\
&\ket{\nu_2 ^2} = \frac{1}{\sqrt{2}} (\overline{\eta} \ket{0} +\eta \ket{2}),
\end{split}
\end{equation}
where $\{\ket{0}, \ket{1}, \ket{2}\}$ denotes the standard basis in $\mathcal{H}$, $\eta = \exp(2/3 \pi i)$, and $\overline{z}$ is used for the complex conjugate of $z$. These vectors allow one to define the measurement operators: $\Pi_{i}^j := 1/3 \ket{\nu_i^j} \bra{\nu_i^j}$, which satisfy all necessary conditions for a SIC-POVM.

The other set of measurement operators is obtained from MUBs, which in the case of qutrits can be represented as \cite{Kurzynski2016}:
\begin{equation}\label{mubs}
\begin{split}
& \ket{\xi_1 ^{(1)}} = \ket{0}, \hspace{0.5cm}  \ket{\xi_2 ^{(1)}}, = \ket{1} \hspace{0.5cm} \ket{\xi_3 ^{(1)}} = \ket{0}, \\
&\ket{\xi_1 ^{(2)}} = \frac{1}{\sqrt{3}} \begin{pmatrix} 1 \\ 1\\1 \end{pmatrix}, \hspace{0.05cm}  \ket{\xi_2 ^{(2)}} =  \frac{1}{\sqrt{3}} \begin{pmatrix} 1 \\ \eta \\ \overline{\eta} \end{pmatrix}, \hspace{0.05cm} \ket{\xi_3 ^{(2)}} =  \frac{1}{\sqrt{3}} \begin{pmatrix} 1 \\ \overline{\eta} \\ \eta \end{pmatrix}, \\
&\ket{\xi_1 ^{(3)}} = \frac{1}{\sqrt{3}} \begin{pmatrix} \eta \\ 1\\1 \end{pmatrix}, \hspace{0.05cm}  \ket{\xi_2 ^{(3)}} =  \frac{1}{\sqrt{3}} \begin{pmatrix} 1 \\ \eta \\ 1 \end{pmatrix}, \hspace{0.05cm} \ket{\xi_3 ^{(3)}} =  \frac{1}{\sqrt{3}} \begin{pmatrix} 1 \\ 1 \\ \eta \end{pmatrix}, \\
&\ket{\xi_1 ^{(4)}} = \frac{1}{\sqrt{3}} \begin{pmatrix} \overline{\eta} \\ 1\\1 \end{pmatrix}, \hspace{0.05cm}  \ket{\xi_2 ^{(4)}} =  \frac{1}{\sqrt{3}} \begin{pmatrix} 1 \\ \overline{\eta} \\ 1 \end{pmatrix}, \hspace{0.05cm} \ket{\xi_3 ^{(4)}} =  \frac{1}{\sqrt{3}} \begin{pmatrix} 1 \\ 1\\ \overline{\eta} \end{pmatrix}.
\end{split}
\end{equation}
The vectors from \eqref{mubs} are used to define projective measurements: $P_i^{(j)} := 1/4 \ket{\xi_i ^{(j)}} \! \bra{\xi_i ^{(j)}}$, which span the space of the linear operators in $\mathcal{H}$ and sum to $\mathbb{1}_3$ (thanks to the normalization constant). By means of the MUBs, we can define $12$ measurement operators which can be considered an informationally overcomplete POVM.

\subsection{Performance analysis}

In our QST framework, we perform qutrit reconstruction by noisy photon counts as introduced in Sec.~\ref{methods} with two sets of measurement operators given in Sec.~\ref{measurements}. To quantify the efficiency of the framework, we introduce three figures of merit.

First, every estimate, $\rho$, obtained by the tomographic technique is compared with the original state produced by the source, i.e. $\ket{\psi_{in}}$ given in \eqref{met4}, by computing quantum fidelity \cite{Jozsa1994,Bengtsson2006}:
\begin{equation}
    F :=  \left(\tr \sqrt{\sqrt{\rho}\ket{\psi_{in}} \!\bra{\psi_{in}} \sqrt{\rho}} \right)^2.
\end{equation}
Then, the average fidelity over the sample is calculated to evaluate the accuracy of the framework. This figure can be treated as a function of the dark count rate and, for this reason, is denoted as $F_{av} (p)$.

Since the framework is restricted to pure states, it is justified to measure the purity of the estimates \cite{Nielsen2000}:
\begin{equation}
    \gamma := \tr\, \rho^2
\end{equation}
and the von Neumann entropy:
\begin{equation}
    S := - \tr (\rho \ln \rho ).
\end{equation}
Analogously as in the case of fidelity, the average purity (entropy) of the estimates is computed over the sample and denoted by $\gamma_{av} (p)$ or $S_{av} (p)$, respectively.

The three figures of merit allow us to study the average performance of the framework versus the amount of noise quantified by $p$. Each figure of merit is computed for a sample of input states along with the standard deviation (SD), which is used to evaluate the amount of variation in the sample.

\section{Results and analysis}\label{results}

\begin{figure}[H]
     \centering
     \begin{subfigure}[b]{\linewidth}
     \caption{}
         \centering
         \includegraphics[width=1\linewidth]{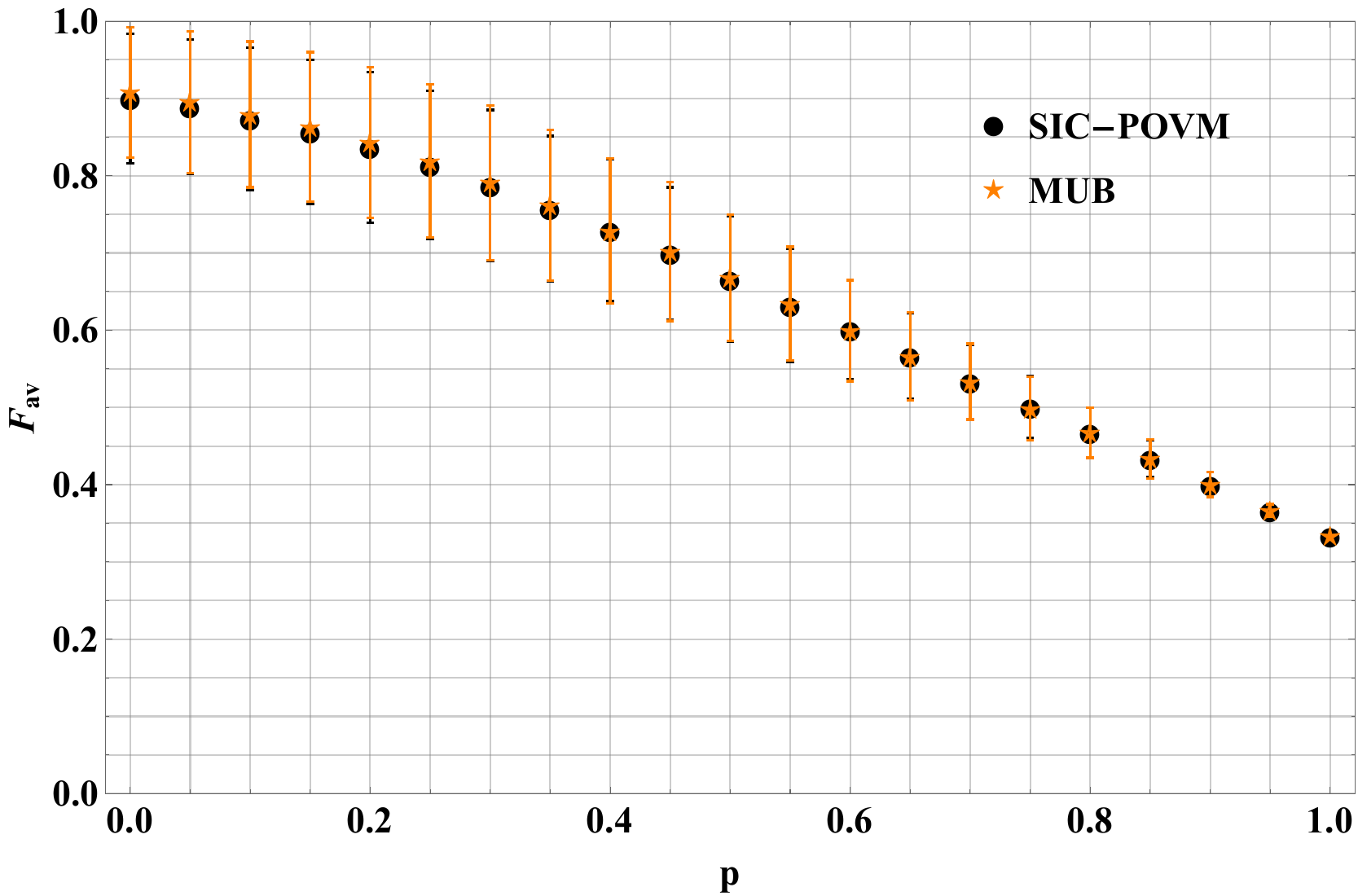}
         \label{fidelity10}
     \end{subfigure}
     \hfill
     \begin{subfigure}[b]{\linewidth}
      \caption{}
         \centering
         \includegraphics[width=1\linewidth]{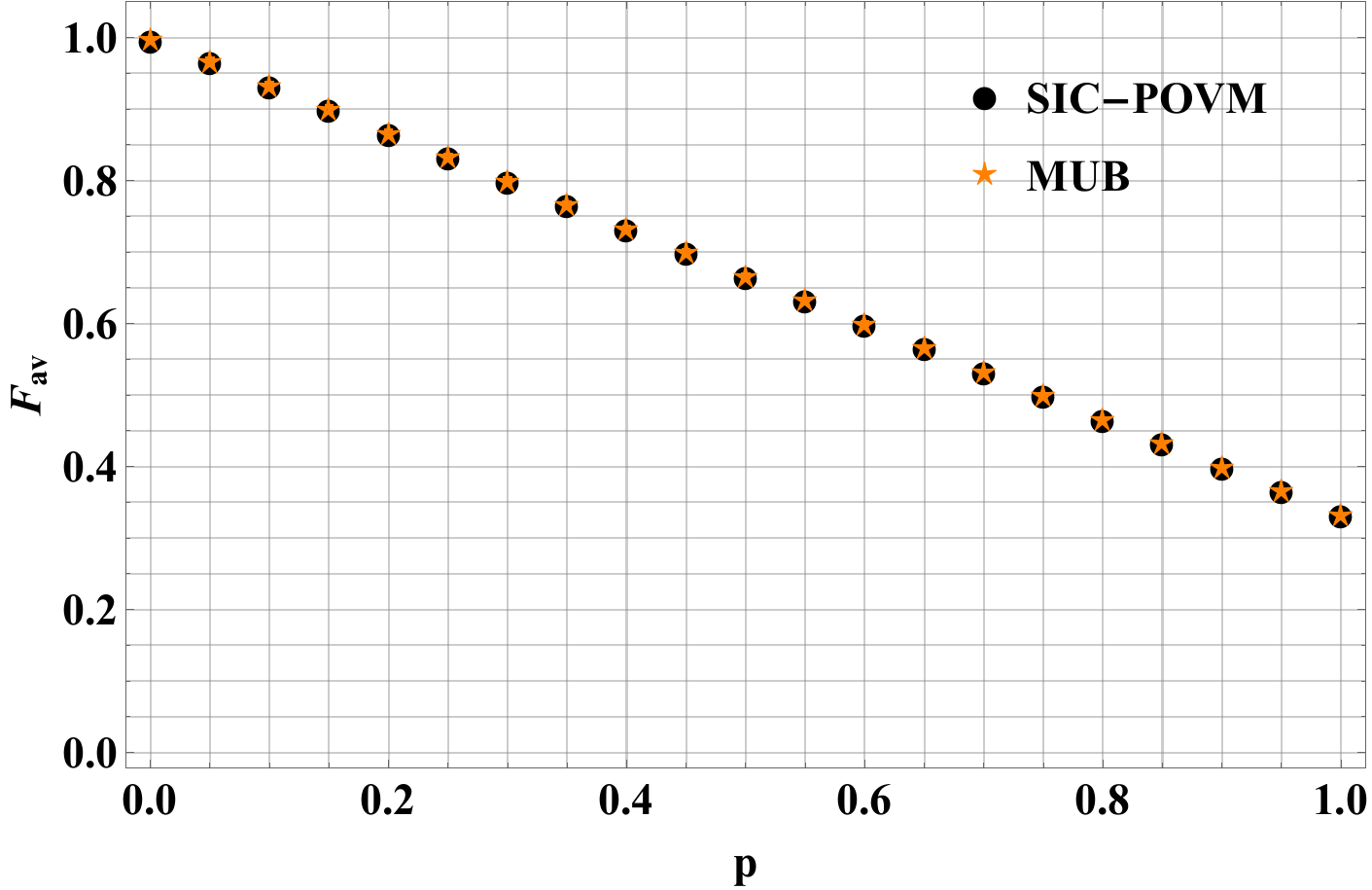}
         \label{fidelity10000}
     \end{subfigure}
        \caption{Plots present the average fidelity $F_{av}(p)$ for the mean number of photons: $\mathcal{N} = 10$ (the upper graph) and $\mathcal{N} = 10\,000$ (the lower graph) in QST of qutrit systems with two measurement schemes: the SIC-POVM and MUBs. Error bars correspond to one SD.}
        \label{figure1}
\end{figure}

First, a sample of $5\,184$ qutrit states was selected according to \eqref{met4}, such that the parameters range over the full scope. Then, each input state goes through the framework, which means that we numerically generate measured photon counts as defined in \eqref{met6}. Next, by the method of least squares \eqref{met7}, we obtain the estimate of our quantum state. This procedure is repeated for each state from the sample, for distinct values of the parameters describing the setup. To be more specific, we consider two different sets of measurements, i.e. the SIC-POVM and MUBs as defined in \ref{measurements}, and two average numbers of photons, i.e. $\mathcal{N} = 10$ (single-photon scenario) and $\mathcal{N} = 10\,000$ (many-photon scenario). The dark count rate is treated as an independent variable.

In \figref{figure1}, one can observe the plots of $F_{av}(p)$ for different measurement settings. One can track how the accuracy of state reconstruction degenerates as we add more dark counts into the scheme. In particular, for $\mathcal{N} = 10$ we notice that even in the absence of dark counts (i.e. $p=0$) we reach the average fidelity $\approx0.9$, see \figref{fidelity10}. This proves that the single-photon scenario is more vulnerable to the Poisson noise. At the same time, we can detect a minor advantage of the MUBs for lower values of the dark count rate (for $p < 0.4$). Especially with no dark counts, we have $F (0) = 0.91 \pm 0.09$ by the MUBs, whereas for the SIC-POVM we obtain $F (0) = 0.90 \pm 0.09$. In such circumstances, the MUBs appear to be more robust against the Poisson noise. However, the difference is not sufficiently significant to announce that the MUBs are a preferable scheme if we utilize few photons per measurement.

By comparing \figref{fidelity10} and \figref{fidelity10000} we see how the accuracy of QST depends on the average number of photons per measurement. By increasing the number of photons up to $10\,000$, we obtain perfect accuracy in the absence of dark counts. To be more specific, we have $F (0) = 0.995 \pm 0.005$ for both measurement schemes. Then, as we boost the dark count rate, the average fidelity declines linearly. In \figref{fidelity10000}, there is no noticeable difference between the measurement schemes.

Furthermore, we can discuss the efficiency of the QST framework in reference to statistical dispersion, which is quantified by the SD and presented as error bars in \figref{fidelity10} and \figref{fidelity10000}. Above all, we notice that the results corresponding to the single-photon scenario feature a great amount of variance. This means that the results for the sample are scattered along a wide interval, which makes it difficult to predict the efficiency of the framework for a given input. However, the value of the SD declines as we increase $p$. On the other hand, the results obtained for $\mathcal{N} = 10\,000$ display little variance, which implies that the average fidelity is a reliable predictor of the accuracy of the framework. Since the many-photon scenario is less vulnerable to the random noise, it provides steady performance, and the results for the sample are not dispersed.

\begin{figure}[H]
     \centering
     \begin{subfigure}[b]{\linewidth}
          \caption{}
         \centering
         \includegraphics[width=1\linewidth]{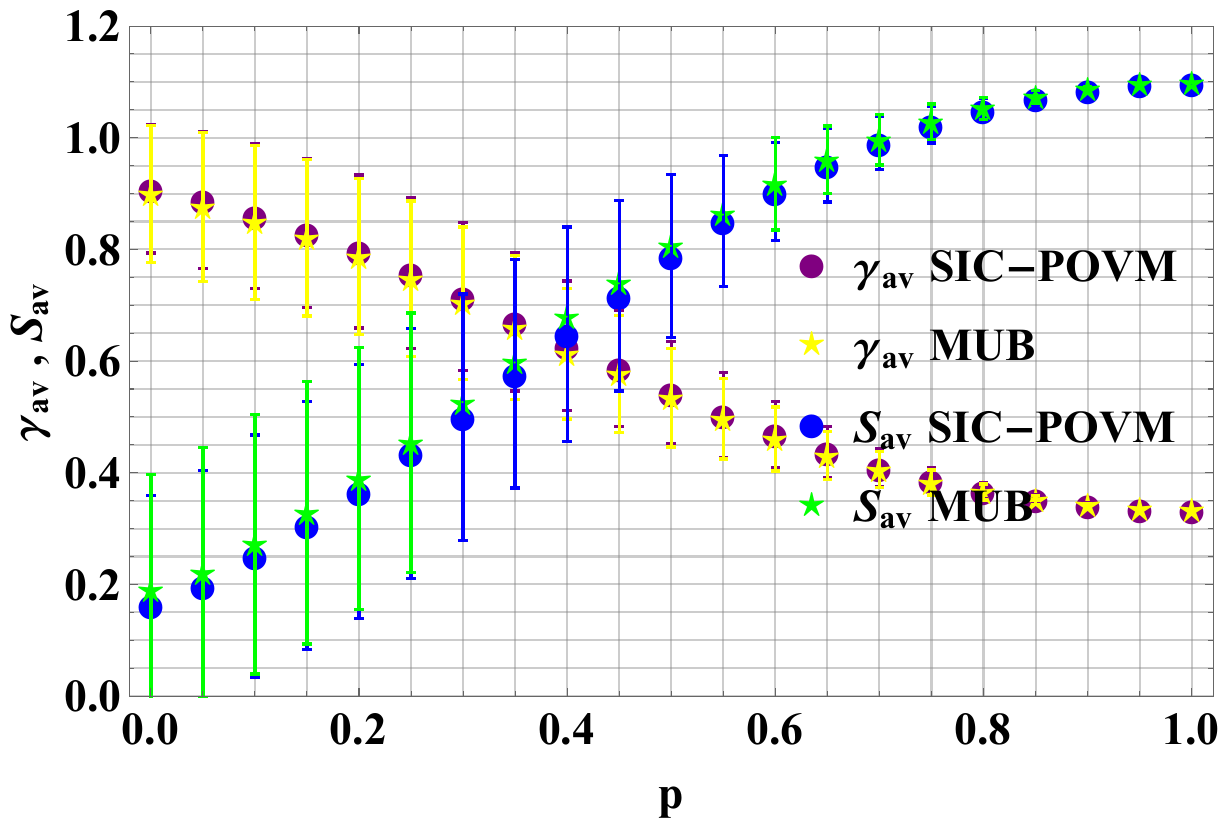}
         \label{purity10}
     \end{subfigure}
     \hfill
     \begin{subfigure}[b]{\linewidth}
          \caption{}
         \centering
         \includegraphics[width=1\linewidth]{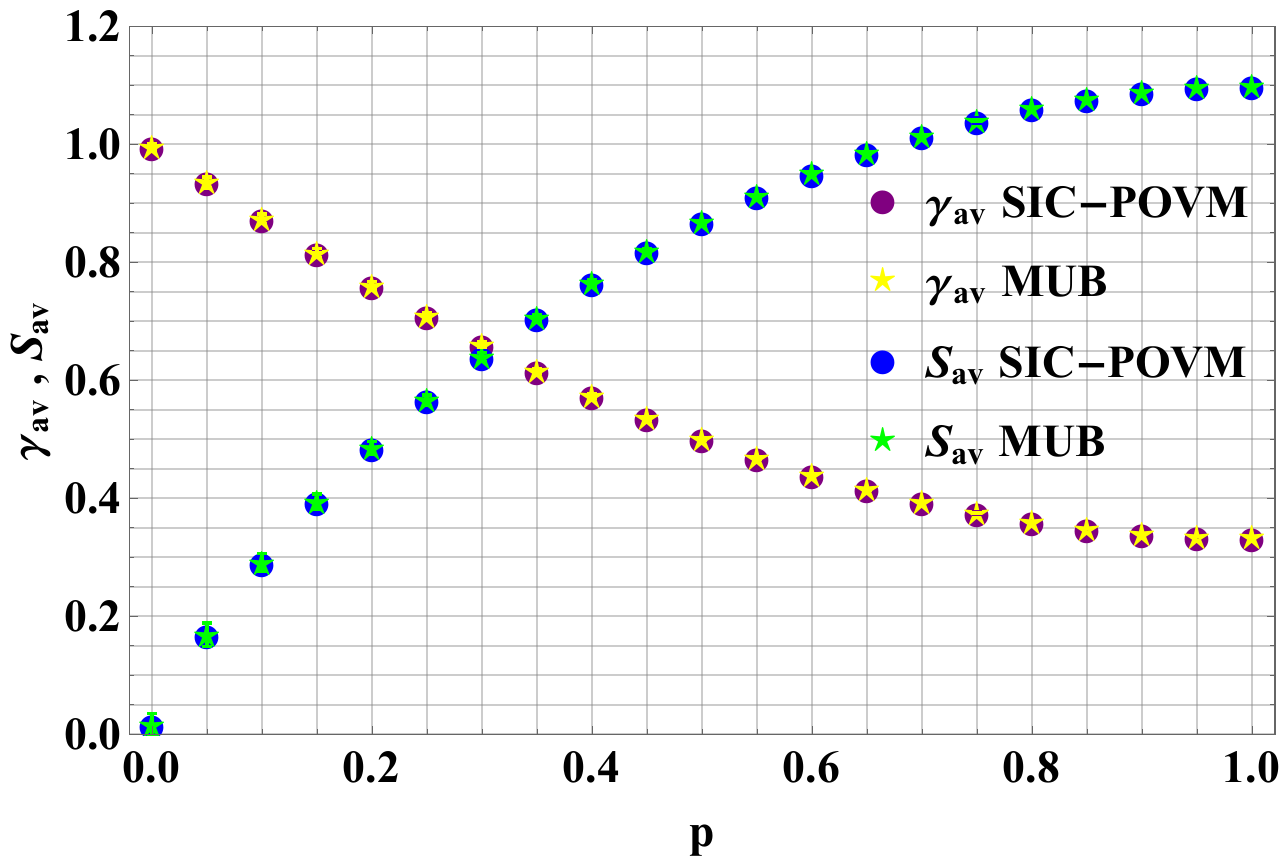}
         \label{purity10000}
     \end{subfigure}
        \caption{Plots present the average purity $\gamma_{av}(p)$ and entropy $S_{av}(p)$ for the mean number of  photons: $\mathcal{N} = 10$ (the upper graph) and $\mathcal{N} = 10\,000$ (the lower graph) in QST of qutrit systems with two measurement schemes: the SIC-POVM and MUBs. Error bars correspond to one SD.}
        \label{figure2}
\end{figure}

In \figref{figure2}, one finds the graphs of the average purity and von Neumann entropy. In \figref{purity10000}, we see that the plots overlap along the entire domain. As we increase the dark count rate, the average purity declines to its minimal value, i.e., $1/3$, whereas the von Neumann entropy grows towards its maximum value, which is $\ln 3$. On the other hand, the plots in \figref{purity10} display some differences when it comes to the measurement schemes. For lower values of the dark count rate, the MUBs lead to quantum states that feature on average more entropy. This result appears intriguing as compared with \figref{fidelity10}, where the measurement based on the MUBs was proved to deliver slightly better accuracy. Nonetheless, just as in the case of fidelity, the results for the single-photon scenario are burdened with significant variance.

\section{Discussion and summary}\label{discussion}

In the article, we studied the problem of quantum state reconstruction of photonic qutrits for different measurement scenarios. We demonstrated how the Poisson noise and dark counts influence the accuracy of state estimation. Two measurement schemes were compared with respect to their efficiency. One was based on the SIC-POVM, whereas the other was defined by the vectors from the MUBs. In particular, we discovered that in the absence of dark counts the measurements with the MUBs slightly outperform the other scheme if the single-photon scenario is considered. Although this observation is in agreement with earlier findings devoted to qubit tomography \cite{Czerwinski2021}, the difference does not appear to be statistically significant.

Additionally, the results provide deeper insight and understanding of the statistical dispersion involved with state estimation. In particular, we discovered that the single-photon scenario features a great amount of variance, which implies that the results for particular states may vary significantly. However, for the many-photon scenario, the results are concentrated close to the mean value, which allows one to precisely predict the accuracy of the framework.

In the future, QST of multilevel quantum systems with noisy measurements will be considered, while special attention will be paid to entangled qutrits. Theoretical research based on numerical simulations can provide valuable insight into the performance of tomographic techniques with imperfect measurements. This knowledge can facilitate future experiments and implementations of quantum protocols since well-characterized quantum resources are required for reliable encoding of information.

\section*{Contributions}

Both authors contributed equally to this work.

\section*{Acknowledgments}

J.~S. was supported by the National Science Centre, Poland (NCN) (Grant No. 2016/23/D/ST2/02064). A.~C. acknowledges financial support from the Foundation for Polish Science (FNP) (project First Team co-financed by the European Union under the European Regional Development Fund).

\end{document}